\documentclass[12pt]{iopart}
\input epsf

\newcommand{\apj}{\textit{Astrophys.\ J.} }
\newcommand{\apjs}{\textit{Astrophys.\ J.\ Suppl.\ Ser.} }
\newcommand{\aap}{\textit{Astron. \& Astrophys.} }
\newcommand{\prb}{\textit{Phys.\ Rev.} B }
\newcommand{\pre}{\textit{Phys.\ Rev.} E }
\newcommand{\pra}{\textit{Phys.\ Rev.} A }
\newcommand{\jpf}{\textit{J.\ Phys.} F }
\newcommand{\jpa}{\textit{J.\ Phys.} A }
\newcommand{\prl}{\textit{Phys.\ Rev. \ Lett.} }

\def\msol{M_\odot}

\def\mjup{M_{\rm J}}

\def\mearth{\,{\rm M}_\oplus}

\def\simgr{\,\hbox{\hbox{$ > $}\kern -0.8em \lower 1.0ex\hbox{$\sim$}}\,}
\def\simle{\,\hbox{\hbox{$ < $}\kern -0.8em \lower 1.0ex\hbox{$\sim$}}\,}

\def\gcc{\,\,\,{\rm g}\,{\rm cm}^{-3}}             
\def\gc2{\,\,\,{\rm g}\,{\rm cm}^{-2}}             

\def\beq{\begin{equation}}
\def\eeq{\end{equation}}

\begin{document}

\title[]{Plasma physics and planetary astrophysics}

\author{Gilles Chabrier}

\address{Ecole Normale Sup\'erieure de Lyon, CRAL, 69364 Lyon Cedex 07, France}
\ead{chabrier@ens-lyon.fr}

\begin{abstract}
In this review, I briefly summarize the present status of experimental and theoretical investigations of the properties of matter under conditions characteristic of planetary interiors, from terrestrial to jovian planets. I first focus on the two lightest elements, hydrogen and helium, and discuss recent theoretical and experimental investigations of their properties at high pressure and temperature. Then, I discuss the impact of these properties,
as well as of the equation of state of heavier elements, on planetary interiors. Finally, I highlight the importance of exoplanet transit observations  and of the inferred mass-radius relationships to determine the planetary interior compositions.
\end{abstract}

\maketitle

\section{Introduction}

Besides our own Solar System jovian planets, more than 400 extrasolar planets (often also referred to as "exoplanets") have been discovered orbiting stars other than our Sun, most of these latter being solar-type stars. The planet masses are determined by the spectroscopic measurement of the motion of absorbing lines (Doppler effect) in the parent star atmosphere induced by the orbital motion of the planet, according to the Kepler's laws. The mass domain covered by these exoplanets ranges from about 10 Jupiter masses (1 $\mjup=1.9\times 10^{27}$ kg $\approx 10^{-3}\msol$) down to about 2 Earth masses (1 $\mearth=5.97\times 10^{24}$ kg $\approx 3\times 10^{-3}\,\mjup$). About 10\% of these extrasolar planets are observed eclipsing their parent star. The star's diminished brightness during the planet's transit provides a direct measure of the ratio of the star and planet surfaces. The star radius being determined by appropriate stellar models, the transit light curve thus enables the determination of the planet's radius, and then of its mean density, providing a strong constraint about the planet's mean composition.

The correct determination of the interior structure and evolution of these planets, and of their genesis, depends on 
the accuracy of the description of the thermodynamic properties of matter under the relevant conditions of temperature and
pressure. These latter reach up to about 10000 K and 10 Mbar (1000 GPa) for Jupiter typical central conditions.
These objects  are composed of an envelope of hydrogen and helium, with some heavier material enrichment, and a core of heavy elements.
The heavier elements
consist of C, N and O, often refered to as "ices" under their molecule-bearing volatile forms (H$_2$O, the most abundant of these elements for
solar C/O and N/O ratios, CH$_4$, NH$_3$, CO, N$_2$ and possibly CO$_2$). The remaining constituents consist of silicates (Mg, Si and O-rich material) and iron (as
mixtures of more refractory elements under the form of metal, oxyde, sulfide or substituting for Mg in
the silicates). 
In the pressure-temperature ($P$-$T$) domain
characteristic of planet interiors, elements 
go from a molecular or atomic state in the low-density outermost regions to an ionized, metallic one in the dense inner parts, covering the regime of {\it pressure}-dissociation and ionization. Interactions between molecules, atoms, ions and electrons are
dominant and degeneracy effects for the electrons play a crucial role, making the derivation of
an accurate equation-of-state (EOS) a challenging task. 
Furthermore, complex phenomena such as phase transition or phase separation may take place in the interior of planets, presenting challenging theoretical problems.

The correct description of the structure and cooling of these astrophysical bodies thus requires the knowledge
of the EOS and the transport properties of the characteristic material under the aforementioned appropriate density and temperatre conditions.
In this short review, we examine our present understanding of the properties of matter under planetary conditions. As will be shown in the next sections, modern experiments and observations
provide stringent constraints on such EOS models.

\section{Hydrogen and helium equation of state}
\label{hug}

\subsection{Hydrogen equation of state}

The quest for the experimental evidence of the pressure-ionization or metallization of hydrogen has remained a challenging problem since the pioneering work of Wigner \& Huntington \cite{WH35}.
A lot of experimental work has been devoted to this problem, but no conclusive result has been reached yet.
Several high-pressure
shock wave experiments have been conducted in order to probe the EOS of deuterium,
the isotope of hydrogen, in the regime of pressure ionization. Gas gun shock compression experiments were
generally limited to pressures below 1 Mbar \cite{Nellis83}, probing only the domain
of molecular hydrogen.
New techniques include laser-driven shock-wave experiments \cite{Collinsetal98,Collinsetal01,Mostovych}, pulse-power compression experiments \cite{Knudsonetal04} and
convergent spherical shock wave experiments \cite{Belovetal02, Boriskovetal03} and
 can achieve pressures
up to 5 Mbar in fluid deuterium at high temperature, exploring
for the first time the regime of pressure-dissociation. These recent experiments
give different results at $P\simgr 1$ Mbar, however, and this controversy needs to be settled before a robust comparison between
experiment and theory can be made in
 the very domain of hydrogen pressure ionization.

On the theoretical front, a lot of effort has been devoted to describing the pressure ionization of hydrogen. The EOS commonly used
for modeling Jovian planet interiors is the
Saumon-Chabrier-Van Horn (SCVH) EOS \cite{C90,SC91,SC92,SCVH} wich includes a detailed description of the partial ionization regime. This EOS
reproduces the Hugoniot data of Nellis \etal 
\cite{Nellis83} but
yields temperatures about 30\% higher than the gas reshock data, indicating insufficient D$_2$
dissociation \cite{Holmes}. A slightly revised version \cite{Saumonetal00} recovers the gas gun reshock temperature data as well as the laser-driven shock wave
results
\cite{Collinsetal98}, with a maximum compression factor of $\rho/\rho_0\simeq 6$,
where $\rho_0=0.17\,\gcc$ is the initial density of liquid deuterium at 20 K.
On the other hand, the earlier
SESAME EOS
\cite{Kerley}, based on a similar formalism, predicts a smaller compression factor, with $\rho/\rho_0\simeq 4$,
in general agreement with all the other recent shock wave experiments. \textit{Ab initio} approaches for the description of dense hydrogen
include path integral Monte Carlo (PIMC)
 \cite{P94,MilitzerCeperley, Militzeretal, Bez04} and
quantum Molecular Dynamics (QMD) simulations.
The latter 
combine molecular dynamics (MD) and Density Functional Theory (DFT) to take into account the quantum nature of the electrons
\cite{Lenosky, Bagnier, Desjarlais, Bonev}. The relevance of earlier MD-DFT calculations was questioned on the basis that these simulations were
unable to reproduce data from gas-gun experiments \cite{Lenosky}. This problem has been solved with more accurate simulations \cite{Bagnier, Desjarlais, Bonev, P04, Delaney06,Vorberger07,MHVTB}. Even at highest densities, not high enough however for hydrogen to behave like an ideal fermion gas, significant
deviations are found between the SCVH EOS and DFT-MD EOS~\cite{MHVTB}.

Although an \textit{ab initio} approach is more satisfactory than the phenomenological approach
based on effective potentials, in practice these simulations
also rely on
 approximations, such as the handling of the so-called
sign problem for the antisymmetrization of the fermion wave functions, or the calculation of the electron functional density itself
(in particular the exchange and correlation effects), or the use of effective pseudo-potentials of restricted validity, not mentioning finite size effects. Moreover, these simulations are too
computationally intensive for the calculation of
an EOS covering several orders of magnitude in density and temperature, as necessary for the description of the structure and evolution of
astrophysical bodies.

Figures 4 and 5 of \cite{SG04} and Figure 1 of \cite{CSP06} compare experimental and theoretical Hugoniots in the $P$-$\rho$ and $P$-$T$ planes.
The disagreement between the laser-driven experiments and the other techniques is clearly illustrated in
the $P$-$\rho$ diagram. Whereas the SCVH EOS achieves a maximum compression similar to the laser-driven data,
all the other models predict compression factors in the $P$-$\rho$ plane in agreement with the more recent data. The MD-DFT results, however,
predict temperatures for the second shock significantly larger than the
experimental results \cite{Holmes}. It is unclear whether this disagreement in the $T$-$V$
plane stems from a significantly underestimated experimental double-shock temperature, due to unquantified thermal conduction
into the window upon shock reflection, or from inaccuracies in the MD-DFT method. As mentioned above, the degree of molecular dissociation, for instance, has a significant influence on the thermodynamic properties
of the fluid and insufficient dissociation in the simulations may result
in overestimates of the temperature. It has
been suggested that the LDA/GGA approximations used in MD-DFT underestimate the dissociation energy of D$_2$ \cite{Stadele}. This would lead to even less dissociation. The fact that compression along the experimental Hugoniot remains small thus suggests compensating effects in the case of hydrogen. More recent,
improved simulations \cite{Bonev}, however, seem to partly solve this discrepancy and to produce reshock temperatures in better
agreement with the experimental results. Peak compression in the modern MD-DFT simulations occurs in the $\sim 0.2$--0.5 Mbar
range around a dissociation fraction of $\sim 50\%$.

These differences in the behaviour of hydrogen at high density and temperature bear important consequences
for the structure and evolution of our Jovian planets.
Jupiter and Saturn are composed by more than 70\%  by mass of hydrogen and helium. Temperatures and pressures range from $T=165$ K and $T=135$ K at $P=1$ bar, respectively, at the surface, to $T>8000$ K, $P>10$ Mbar at the center. At pressures around $P\sim 1$--3 Mbar, corresponding to about $80\%$ and $60\%$ of the planet's radius, as measured from the planet center,  for Jupiter and Saturn, respectively, hydrogen thus undergoes a transition from an insulating molecular phase to a conducting ionized plasma. 
The differences in the hydrogen Hugoniot experiments must be correctly understood
before the description of hydrogen pressure dissociation and ionization stands on firm grounds.
As noted by Boriskov \etal \cite{Boriskov}, all the recent experiments agree quite well in terms of the shock speed $u_s$ versus the particle velocity $u_p$, almost within their respective error bars.  Error bars and differences in $(u_s,u_p)$ are amplified
in a $P$--$\rho$ diagram by a factor of  $({\rho / \rho_0}-1)$, due to the Rankine-Hugoniot shock conditions.  These are challenging experiments as the differences highlighted in
panel 1 of Fig.~1 of \cite{CSP06} arise from differences in $u_s$ and $u_p$ of less than 3\%. High-pressure isentropic compression experiments, planed for a near future, are promising techniques to help address this challenge.

\subsection{Helium equation of state}

The planet interior models are also affected, to a lesser extent, by the uncertainties of the helium EOS. A model EOS for
helium at high density, covering the regime of pressure ionization and improving upon the previous description of dense helium in SCVH, has been developed recently by Winisdoerffer and Chabrier
\cite{WC05}. This EOS, based on effective interaction potentials between He, He$^+$, He$^{++}$ and e$^-$ species, 
adequately reproduces experimental Hugoniot and sound speed measurements up to $\sim 1$ Mbar. In this model,
pressure ionization is predicted to occur directly from He to He$^{++}$ for $T\simle 10^5$ K. Because of the uncertainties in the treatment of the
interactions at high density, however, the predicted ionization density ranges from a few to $\sim 10 \,\gcc$, i.e. $P\sim 9-20$ Mbar, depending on the temperature. This is ignificantly larger that the $\rho \sim 1 \gcc$ density above which available measurements of electrical conductivity of helium predict that the plasma is substantially ionized \cite{Ternovoi01, Fortov03}. These measurements, however, conflict with MD-DFT conductivity calculations \cite{Kowalski07}. It must be kept in mind that the reported measurements are model dependent and that the conductivity determinations imply some underlying EOS model. PIMC and DFT-MD simulations have also been applied to
helium~\cite{Mi06,StixrudeJeanloz,Mi09}.
 Recent high-pressure experiments, using statically precompressed samples in dynamical compression experiments, have been achieved up to 2 Mbar for various Hugoniot initial conditions, allowing to test the EOS over a relatively broad range \cite{Eggert08}. These experiments show a larger compressibility than for hydrogen, due to electronic excitations, and are
in good agreement with the SCVH EOS while ab-initio calculations \cite{Mi06} tend to underestimate the compressibility \cite{Eggert08}. Clearly more of these experiments, exploring a higher pressure range to reach the helium pressure {\it ionization} regime, are needed to fully assess the validity of the various EOS models, with an important impact on our knowledge of the structure of Jovian planets.

\subsection{Hydrogen metallization}

As mentioned earlier, the nature of the metallization of hydrogen remains an open issue, of prime importance for giant planet structure, evolution and magnetic field generation. Several calculations based on the free energy minimization method \cite{Norman,Ebeling,SC89, SC92} predict that hydrogen pressure ionization should occur through a first-order transition, the so-called plasma phase transition (PPT). Nearly all these PPT calculations are based on a model Helmholtz free energy that includes contributions from (i) neutral particles (H and H$_2$), (ii) charged particles (protons and electrons), and usually (iii) some coupling interaction between these species. It is well known that fully ionized plasma models become thermodynamically unstable (negative specific heat or isothermal compressibility) at low temperature and moderate densities. This is analogous to the behaviour of expanded metals at $T=0$ that display a region where $dP/d\rho<0$ \cite{PinesNozieres}. This behaviour of the fully ionized plasma reflects the formation of bound states in a real system but is formally a flaw of the model. In other words, even though a first order transition might be real, it is built by construction in all aforementioned free energy based models and makes the PPT prediction from these models not credible. First principle methods, on the other hand, yield different predictions. Some calculations \cite{Scandalo03,Bonev} find a sharp (6$\pm$2\%) volume discontinuity at constant pressure or $dP/dT<0$ at constant volume \cite{Magro96}, a feature consistent with a first order transition. At the same time, the pair correlation function exhibits a drastic change from a molecular to an atomic state with a metallic character (finite density of electronic states at the Fermi level). These transitions are found to occur in the $\sim 0.5$-$1.25$ Mbar and $\sim 1500$-3000 K domain.
Note that a first order structural transition at $T=0$ is predicted to occur at a pressure $P\simgr 4.0$ Mbar, from DFT calculations based on exact exchange calculations \cite{Stadele}.
However, more recent calculations, based on a more accurate Born-Oppenheimer wave function propagation method than the aforementioned previous studies, find a gradual, continuous transition from hydrogen insulating to conducting state at high density ~\cite{Bagnier,Delaney06,Vorberger07}, although eventually with a region of
$\partial P / \partial T|_V<0$. On the experimental front, the question remains unsettled. Recent shock wave
experiments~\cite{Fortov07} show evidence of an abrupt insulator-to-metal
transition at temperatures and pressures consistent
with theoretical predictions, identified by the authors as the signature of a PPT. Given the experimental error bars, however, these results must be considered with caution and need to be
confirmed or rejected by further experiments before conclusions about the very nature of hydrogen pressure ionization/metalization can be considered as robust.

\subsection{Hydrogen-Helium phase separation}

The existence of a phase separation between hydrogen and helium under conditions characteristic of Jupiter and Saturn interiors was first suggested independently by
Smoluchowski \cite{Smoluchowski73} and Salpeter \cite{Salpeter73} and the first detailed calculations were conducted by Stevenson \& Salpeter \cite{SS77}. A phase separation is a first order transition which implies a concentration and thus a density discontinuity below a critical temperature, as
given by the condition:
\beq
\mu_i^I=\mu_i^{II}\Rightarrow x_i^I=x_i^{II}\,e^{-\frac{\Delta G}{kT}},
\eeq
where $\mu_i^I,x_i^I$ denote respectively the equilibrium chemical potential and number concentration of the species $i$ in phase $I$ (resp. in phase $II$),
 and $\Delta G$ is the excess (non ideal) mixing enthalpy between the two phases.
 Under the action of the planet's gravity field, a density discontinuity yields an extra source of gravitational energy as the dense phase dropplets (namely He-rich
 ones in the present context) sink towards the planet's center. Conversion of this gravitational energy into heat delays the cooling of the planet,
 which implies a larger age to reach a given luminosity compared with a planet with a homogeneous interior (conversely, the planet appears younger for its observed luminosity than would
 the same object without this extra source of energy). The time delay ican be estimated as:
 \beq
 \Delta t\approx \frac{\Delta E}{L}\approx {(\Delta M)g\frac{\Delta \rho}{\rho} {\bar R}\over L},
 \eeq
 where $\Delta M$ is the mass fraction experiencing phase separation, $\Delta \rho$ is the density difference between the two phases, $g$ is the planet's gravity, $ {\bar R}$ the planet's mean radius ($\approx R/2$) and $L(T)$ its luminosity at temperature $T$.
 
 In Saturn's case, such an additional source of energy is required to explain the otherwise too bright luminosity at the correct age, i.e. the age of the Solar System, $\sim 4.5\times 10^9$ yr \cite{SS77}. There has been few studies of the H/He phase diagram. Some calculations
 \cite{Stevenson75,HdW85,GC} assume that the phase separation takes place in the fully ionized part of the planet interior, i.e. in the H$^+/$He$^{2+}$ domain. The critical points obtained with these calculations are in the range $T\sim 7000$-10000 K and $P\sim 2$-8 Mbar. The extra release of gravitational energy predicted by these diagrams, however, is found to be insufficient to reconcile Saturn's luminosity with the age of the Solar System \cite{FortneyHubbard04}. DFT electronic structure calculations for the $T=0$ H/He solid alloy, with no assumption on the degree of ionization, were first conducted by Klepeis et al. \cite{Klepeis91}. Finite temperature results were obtained by applying an estimated entropy correction. The critical point is predicted to occur at an unrealisticaly high temperature ($\sim 40000$ K at 10.5 Mbar) and is excluded by the constraints arising from Jupiter and Saturn evolution \cite{Guillot95}. The following, finite temperature MD-DFT calculations, were performed by Pfaffenzeller et al. \cite{Pfaffenzeller95}. The
 predicted critical
 temperature ($\sim 4000$ at 10.5 Mbar) implies that no phase separation should have occured yet in Jupiter and Saturn interiors. The major problem of this work, however, is that it does not recover the fully ionized limit. A striking result of these latter calculations is the prediction of an increasing critical temperature with {\it increasing} pressure, a result qualitatively opposite to the ones obtained with the aforementioned calculations. A major consequence for jovian planet evolution is that, once separation occurs, it will occur within almost the entire planet's interior \cite{Guillot95,FortneyHubbard04}. More recently, PIMC~\cite{Mi05} and DFT-MD ~\cite{Vorberger07} simulations have been applied to H/He mixtures.
Unfortunately, given the aforementioned difficulty in modeling the
properties of H or He alone, and the necessity to simulate a large enough number of particles for
the minor species (10\% by number for He in the present case) to obtain
statistically converged results and a reliable phase diagram, it seems fair and cautious to say that no reliable calculation of the H/He phase diagram can be claimed so far.
The recent claim \cite{StixrudeJeanloz} that H/He phase separation can not take place in Jupiter's interior, because metallization of He should occur at lower pressure than previously expected, is not correct and is based on a misunderstanding. Although such a facilitated metallization (by itself a model-dependent result)
could exclude the suggested H$^+$-He immiscibility \cite{Stevenson79}, it does not preclude the H$^+$-He$^{++}$ one.

\section{Equation of state for heavy elements.}
\label{eosz}

As mentioned in the introduction, the composition of gaseous planets also includes heavier elements under the form of ices, silicates, iron or other compounds.
The behaviour of these different elements as a function of pressure, under the conditions typical of giant planet interiors
is not or poorly known. At
very high pressure, the categorizations of gas, ice and rock becomes meaningless and these elements
should become a mixture of closed-shell ions. The most
widely used EOS models for such elements are ANEOS \cite{ThomsonLauson72} and SESAME \cite{LyonJohnson92}, which describe the thermodynamic properties of water, "rocks" (olivine (fosterite Mg$_2$SiO$_4$) or dunite in ANEOS, a mixture of silicates and other heavy elements called "drysand" in SESAME) and iron. These EOS consist of
interpolations between models calibrated on existing Hugoniot data, with thermal corrections approximated by a Gruneisen parameter ($\gamma=\frac{V}{C_V}(\frac{dP}{dV}_V$)), at low to moderately high
($\simle 0.5$ Mbar) pressure, and Thomas-Fermi or more sophisticated first-principle calculations at very
high density ($P\simgr 100$ Mbar), where ionized species dominate. Interpolation between these limits, however, provides no insight about the correct structural and electronic properties of
the element as a function of pressure, and thus no information about its compressibility, ionization stage (thus conductibility), or even its
phase change, solid or liquid. All these properties can have a large impact on
the internal structure and the evolution of the planets. Current diamond anvil cell experiments reach several thousands degrees at a maximum pressure of about 2 Mbar for iron \cite{Boehler93}, still insufficient to explore the melting curve at the Earth inner core boundary ($\sim 3$ Mbar and $\sim$ 5000 K). On the other hand, dynamic experiments yield too high temperatures to explore the relevant $P$-$T$ domain for the Earth but may be very useful to
probe e.g. Neptune-like exoplanet interior conditions. As for the phase diagram of water, it has been
explored only up to 0.35 Mbar and 1040 K \cite{Lin05} and the melting curve of water at higher pressure and temperature, typical of icy (Neptune-like) or gaseous (Jupiter-like) giant planet interiors, is presently undetermined.

A detailed comparison between various EOS, including ANEOS and SESAME, for heavy elements, and the impact of the related uncertainties on the planet's radius for Earth-like to Jupiter-like planets has been conducted in \cite{baraffe08}. The largest difference between the various EOS models, reaching up to $\sim 40$-60$\%$ in $P(\rho)$ and $\sim 10$-15$\%$ on the entropy $S(P,T)$,
occurs in the $T\sim10^3$-$10^4$ K, $P\sim10^{-2}$-$1$ Mbar interpolated region, the typical
domain of Neptune-like planets. For these objects, such an uncertainty on the heavy element EOS translates into a $\sim 10\%$ uncertainty in the radius after 1 Gyr,
and to larger uncertainties at earlier ages (see Fig. 3 of \cite{baraffe08}) and prevents precise determinations of the planet internal composition, a key issue for our understanding of planet formation.

\section{Consequences for planetary interiors}
\subsection{Jupiter and Saturn}

The rapid rotation of Jovian planets induces a nonspherical gravitational field that can be expanded 
in Legendre polynomials $P_n(\cos \theta)$:
\begin{eqnarray}
V(r,\theta)=-{GM\over r}\Bigl[1-\sum_{n=1}^\infty \left({R_{eq}\over r}\right)^nJ_nP_n(\cos \theta)\Bigr],
\end{eqnarray}
where $M$ and $R_{eq}$ denote respectively the planet's mass and equatorial radius, 
and the $J_n$ are the gravitational moments:
\begin{eqnarray}
J_n=-{1\over MR_{eq}^n}\int_Vr^{\prime n}P_n(\cos \theta)\rho(r^\prime,\theta)\,d^3r^\prime .
\end{eqnarray}

Because of north-south symmetry, the moments of odd order are null. The first three nonvanishing moments, $J_2$, $J_4$ and $J_6$
have been measured with high accuracy for both planets during spacecraft flyby missions. Combined with the planet's mass, radius
and rotation period, these provide integral constraints on the density profile  of the planet, $\rho(r)$, to be compared with the
corresponding values from a structure model
obtained for a self-gravitating and rotating fluid body in hydrostatic equilibrium. The EOS provides the $P(\rho)$
relation needed to close the system of equations. The structure of the H/He envelopes of giant planets is fixed by the specific entropy
determined from observations at their surface. The very high efficiency of convection in the interior of these objects
leads to nearly adiabatic interior profiles.
The structure of the planet is thus primarily determined by the choice of the hydrogen and to a lesser
extent by the helium and heavier element EOS used in the models.
A detailed study of the influence of the EOS of hydrogen on the structure and evolution
of Jupiter and Saturn has been conducted in \cite{SG04}.
Fortunately, some shock wave experiments overlap
Jupiter's and Saturn's adiabats. As demonstrated in this study, the small
($\le 5\%$) difference on the $(P,\rho)$ relation along the adiabat between the various EOSs representative of the two sets of
experimental results on the H EOS at high density (see \S 2.1), is large enough to affect appreciably the interior structure of the models. A slightly modified SESAME EOS, which recovers the H$_2$ entropy at low temperature and density,
yields Jupiter models with a very small core mass, $M_\mathrm{core}\sim 1\,\mearth$ and a mass $M_Z\sim 33\,\mearth$ of heavy elements ($Z>2$)
mixed in the H/He envelope. The SCVH EOS yields models with $M_\mathrm{core}\sim$0--6 $\mearth$ and $M_Z\sim 15$-26$\,\mearth$.
Models of Saturn are 
less sensitive to the EOS differences, since only $\sim 70\%$ of its mass lies at $P>1$ Mbar, compared to 91\%
for Jupiter.
Models computed with the SCVH and the modified SESAME EOS have $M_{\rm core}=10$--21$\,\mearth$ and $M _Z=20$--27$\,\mearth$
and 16--29$\,\mearth$, respectively.
As shown in this study, the temperature along the adiabat is more sensitive to the choice of the EOS. This affects
the thermal energy content of the planet and thus its cooling rate and evolution.
EOS which are adjusted to fit the deuterium reshock temperature measurements \cite{Ross98} lead to models that take
$\sim 3\,$Gyr for Jupiter to cool to its present state, a clearly excluded solution.
Even when considering uncertainties in the models,
or considering the possibility of a  H/He phase separation, 
such
a short cooling age is unlikely to be reconciled with the age of the solar system.
This astrophysical constraint suggests that
the reshock temperature data are too low.

Recently, new models for Jupiter have been derived, both based on the previously mentioned DFT-MD calculations for the H and He EOS. The first models \cite{MHVTB} predict a large, 14-18 $\mearth$ core of heavy material. Since these authors assume a homogeneous interior throughout the entire gaseous H/He envelope, the free condition left to match all the observable constraints (gravitational moments) is the inner rotation profile of the planet. Therefore, these Jupiter models are predicted to have a finitie differential rotation along cylinders, a signature potentially observable by future orbiter missions like {\it JUNO}. On the other hand, the second type of models \cite{Nettelmann08} assume two distinct gaseous envelopes, namely an outer heavy-element depleted molecular region and a deeper heavy-element enriched ionized region, the composition discontinuity between the two regions providing a free parameter to calculate models matching the observational constraints. These models also predict a significant heavy-material enrichment within Jupiter, although with a different inner profile and a much more modest ($\simle 3\mearth$) central core. Note also that the two model interior temperature profiles differ substantially, with the second ones being hotter than the first ones, which implies different cooling histories for the planet.

\subsection{Transiting extrasolar planets}

As mentioned in \S3, current uncertainties in available EOS models for H, He and heavy elements prevent so far accurate determinations of the transiting planet interior compositions. Planets below about 10 $\mearth$, covering the range of Earth-like to "Super Earth" planets,  however, are less affected by these uncertainties. Indeed, these objects are not massive enough to retain a substantial gaseous atmosphere and are composed primarily of refractory elements, ices, rock and iron. The mass-radius relationship for these terrestrial planets has been parametrized as $R=R_{ref}(M/\mearth)^\beta$, with $R_{ref}=(1+0.56\,\alpha)R_\oplus$ and
$\beta=0.262(1-0.138\,\alpha$), for the rocky or ocean Super-Earth planets \cite{valencia07}, where $\alpha$ denotes the water mass fraction, and $\beta=0.3$ for planets between 10$^{-2}$ to 1 $\mearth$,
with a weak dependence upon the iron to silicate ratio Fe/Si \cite{sotin07}. Note that incompressible (constant density)  material corresponds to $\beta=1/3$. These parametrizations appear to be rather robust, despite the uncertainties in the heavy element EOS and in the iron/silicate fraction
\cite{fortney07,seager07,sotin07}. 
Current uncertainties in
the high-pressure behaviour of silicates, ices and iron alloys, however, prevent a more precise determination of
the internal composition or of the size and nature, solid or liquid, of the central core.

\section{Conclusion}
In this brief review, we have considered the description of the thermodynamic properties of dense matter under the specific
conditions of planetary interiors, from Earth-like to Jupiter-like objects. The description of the pressure ionization of hydrogen, helium and other elements,
as well as the possible immiscibility of H and He under these planetary conditions, play an important role in determining the mechanical and thermal properties and the evolution of these objects.
EOS models are still hampered by significant uncertainties but modern ongoing or future experiments and/or observations can enable
us to eventually discriminate between these models.
New experimental set ups, like the LIL ("Ligne d'Integration Laser") in France or the NIF ("National Ignition Facility") in the US should enable us to probe the EOS of various light or heavy elements under conditions previously out of reach, characteristic of the aforementioned planetary interior conditions. At the same time, rapid progress in computer performances should enable us to derive accurate EOS based on first-principle methods. Combined with the wealth of data of transiting extrasolar planets expected from the COROT, KEPLER and future missions, our knowledge of the properties of matter under planetary interior conditions should drastically improve within the near future.


\end{document}